\documentclass[epj]{webofc}
\usepackage[varg]{txfonts}   
\woctitle{QCD@Work 2014}

\begin{document}

\title{Measurement of azimuthal correlations between D mesons and charged hadrons with ALICE at the LHC}

\author{Fabio Colamaria\inst{1,2}\fnsep\thanks{\email{fabio.colamaria@ba.infn.it}} on behalf of the ALICE Collaboration}

\institute{University of Bari ``Aldo Moro'' \and INFN - Section of Bari}

\abstract{
The comparison of angular correlations between charmed mesons and charged hadrons produced in pp, p-Pb and Pb-Pb collisions can give insight into the mechanisms through which charm quarks lose energy in a QGP medium, produced in ultra-relativistic heavy-ion collisions, and can help to recognize possible modifications of their hadronization induced by the presence of the QGP. The analysis of pp and p-Pb data and the comparison with predictions from pQCD calculations, besides constituting the necessary reference for interpreting Pb-Pb data, can provide relevant information on charm production and fragmentation processes. In addition, possible differences between the results from pp and p-Pb collisions can give information on the presence of cold nuclear matter effects, affecting the charm production and hadronization in the latter collision system.

A study of azimuthal correlations between D$^0$, D$^+$, and D$^{*+}$ mesons and charged hadrons in pp collisions at $\sqrt{s}$ = 7 TeV and p-Pb collisions at $\sqrt{s_{\rm NN}}$ = 5.02 TeV are presented. D mesons were reconstructed from their hadronic decays at central rapidity in the transverse-momentum range $3 \leq p_{\rm T}^{\rm D} \leq 16$ GeV/$c$ and were correlated to charged particles reconstructed in the pseudorapidity range $|\eta| < 0.8$. Perspectives for the measurement in Pb-Pb collisions at $\sqrt{s_{\rm NN}}$ = 2.76 TeV will also be presented.
}
\maketitle
\section{Physics motivations}
\label{sec:intro}
ALICE~\cite{bib:ALICE} has measured $p_{\rm T}$-differential cross sections for D-meson production at central rapidity in pp~\cite{bib:Dmes_pp_7, bib:Ds_pp_7, bib:Dmes_pp_2.76} and Pb-Pb~\cite{bib:Dmes_PbPb} collisions. A suppression of D-meson yields of a factor 4$-$5 was found in central Pb-Pb collisions for $p_{\rm T} >$ 5 GeV/$c$ with respect to cross sections measured in pp collisions and scaled by the nuclear overlap function. This feature can be attributed, at least partially, to the energy loss of charm quarks in the Quark Gluon Plasma formed in such collisions. Further insight into this topic can be provided by the study of angular correlations between D mesons and unidentified charged hadrons in Pb-Pb collisions, where medium-induced modifications to charm fragmentation and hadronization and the dependence of the energy loss on the path length in the medium can be addressed.

Such a study is of great interest also in pp and p-Pb collisions; in the former system it allows to characterize charm quark jets and study their properties, besides providing a reference for the measurements in p-Pb and Pb-Pb collisions, while in the latter it enables the study of cold nuclear matter effects and the search for long-range ridge-like structures in near and away side regions (``double ridge''), already observed in hadron-hadron correlations~\cite{bib:DoubleRidge}, in the heavy-flavour sector.

\section{Description of the analysis}
\label{sec:descr}
The analysis was performed for D$^0$, D$^+$ and D$^{*+}$ mesons on a minimum bias sample of $3.1 \cdot 10^8$ pp collisions at $\sqrt{s} = 7$ TeV and on a minimum bias sample of $1.0 \cdot 10^8$ p-Pb collisions at $\sqrt{s_{\rm NN}} = 5.02$ TeV, for different ranges of the D-meson $p_{\rm T}$ and thresholds of the charged hadron $p_{\rm T}$.
D-meson candidates (\textit{trigger} particles) were reconstructed from their hadronic decay channels (D$^0\rightarrow$ K$^-\pi^+$, D$^+\rightarrow$ K$^-\pi^+\pi^+$ and D$^{*+}\rightarrow$ D$^0 \pi^+\rightarrow$ K$^-\pi^+\pi^+$) and selected exploiting their displaced decay vertex topology, PID and reconstruction quality cuts on the daughter tracks. The selected D mesons in the signal region ($\pm 2\sigma$ from the centre of the signal peak in the invariant mass distribution) were then correlated with charged hadrons (\textit{associated tracks}) with $|\eta| < 0.8$ and satisfying a selection on their reconstruction quality, and the difference in pseudorapidity $\Delta\eta$ and azimuthal angle $\Delta\varphi$ between the D-meson candidate and the charged-particle track was evaluated.

The contribution of background candidates was subtracted using the correlation distribution from the candidates in the sidebands of the invariant mass distribution (the regions 4$-$8$\sigma$ away from the signal peak), normalized by the ratio of background candidates in signal and sideband regions. Several corrections were applied to the correlation distributions to account for:

\begin{itemize}
\itemsep-0.3em
\item \textbf{Associated track reconstruction efficiency}. Each D-hadron correlation was weighted by the inverse of the track reconstruction efficiency of the associated track, evaluated from a MC simulation as a function of the primary vertex $z$ coordinate and of the track $p_{\rm T}$ and $\eta$ (while the $\varphi$ distribution of the associated track sample was found to be isotropic);
\item \textbf{D-meson selection efficiency}. Each correlation entry was weighted by the inverse of the D-meson reconstruction and selection efficiency, computed as a function of the D-meson $p_{\rm T}$ and of the event multiplicity from a MC simulation; other minor dependencies were neglected to avoid introducing fluctuations in the efficiency, due to the limited statistics of the MC sample;
\item \textbf{Limited detector acceptance and detector spatial inhomogeneities}. A correction factor was obtained exploiting the event mixing technique, i.e. correlating D-meson candidates from an event with tracks from other events with similar multiplicity and $z$ of the primary vertex, and normalizing the resulting correlation distribution to its value in $(\Delta\varphi$,$\Delta\eta) = (0$,$0)$;
\item \textbf{Beauty feed-down}. The fraction of D mesons originating from B-meson decays was estimated using the reconstruction efficiencies of secondary D mesons and the $p_{\rm T}$-differential cross-section of the latter, obtained using FONLL calculations and the EvtGen package, as described in~\cite{bib:Dmes_pp_7}. MC simulations based on PYTHIA were exploited to evaluate a template distribution of angular correlations between D mesons from B-meson decays and charged hadrons; this distribution was scaled to match the fraction of feed-down over prompt D mesons, calculated as explained above, and subtracted from the inclusive D-hadron correlation distribution obtained from data.
\item \textbf{Secondary track contamination}, i.e. tracks from strange-hadron decays or produced in interaction of particles with the detector material. From a MC study, no visible angular dependence was found for this contamination. Hence a global scale factor, corresponding to the fraction of primary particles in the sample of tracks correlated with D mesons, was applied to the correlation distributions.
\end{itemize}

Due to the limited statistics available, the fully corrected 2D correlations were projected onto the $\Delta\varphi$ axis, producing azimuthal correlation plots, which were normalized to the number of trigger D mesons. A weighted average of the results for the three meson species was then performed.
A fit function, composed of a constant term plus two Gaussian functions with a periodicity condition, was applied to the azimuthal correlation distributions. This allowed to extract physical observables (near-side yield, near-side peak width, height of the baseline of the distribution) to be compared between the two collision systems and with predictions from Monte Carlo simulations. Finally, a careful evaluation of all the sources of systematic uncertainties was performed.

\section{Results in pp and p-Pb collision systems}
\label{sec:results}
\begin{figure}[!]
\centering
\begin{minipage}{\linewidth}
  \centering
  $\vcenter{\hbox{\includegraphics[width=.41\linewidth]{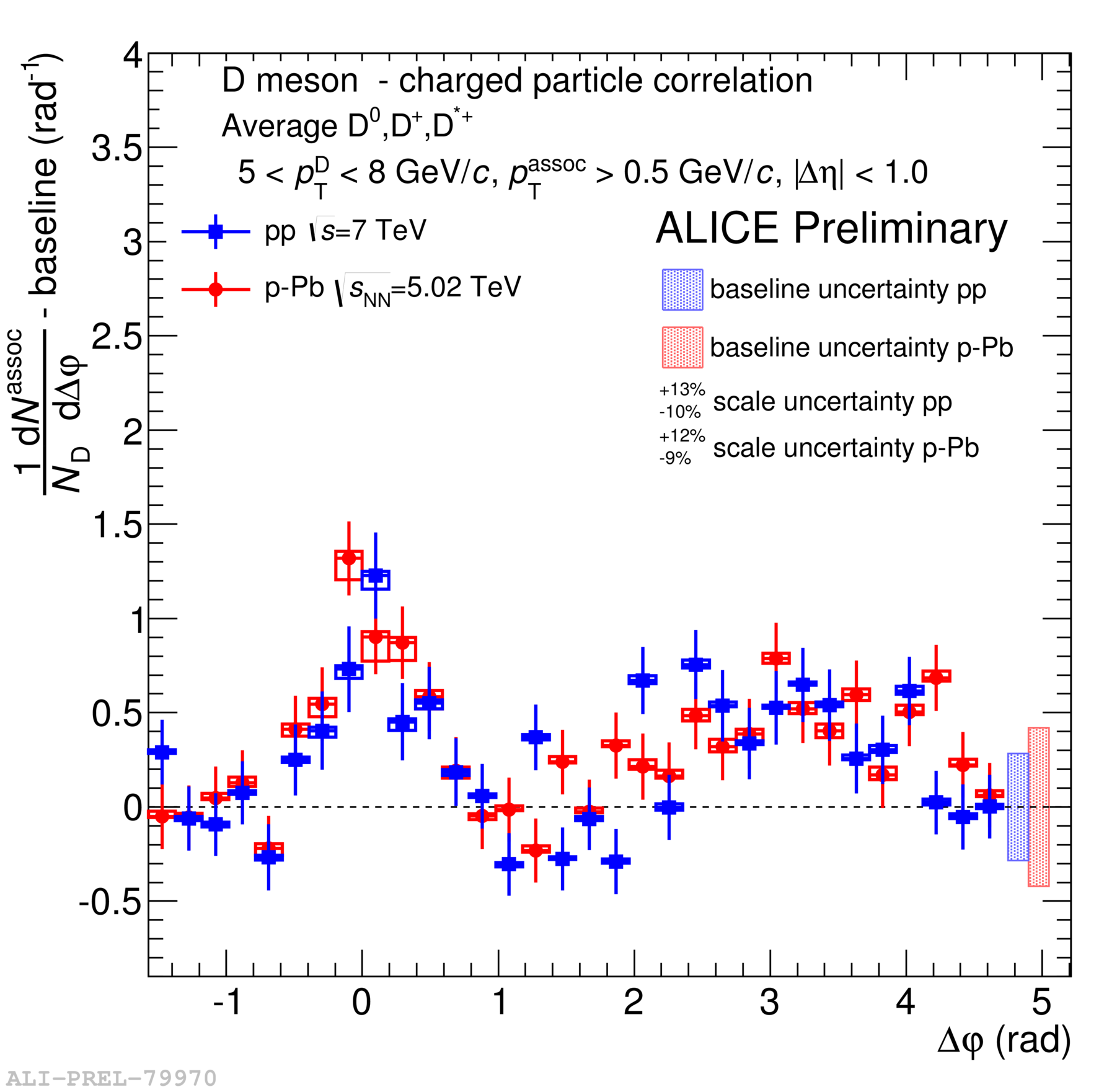}}}$ \hspace{0.5cm}
  $\vcenter{\hbox{\includegraphics[width=.41\linewidth]{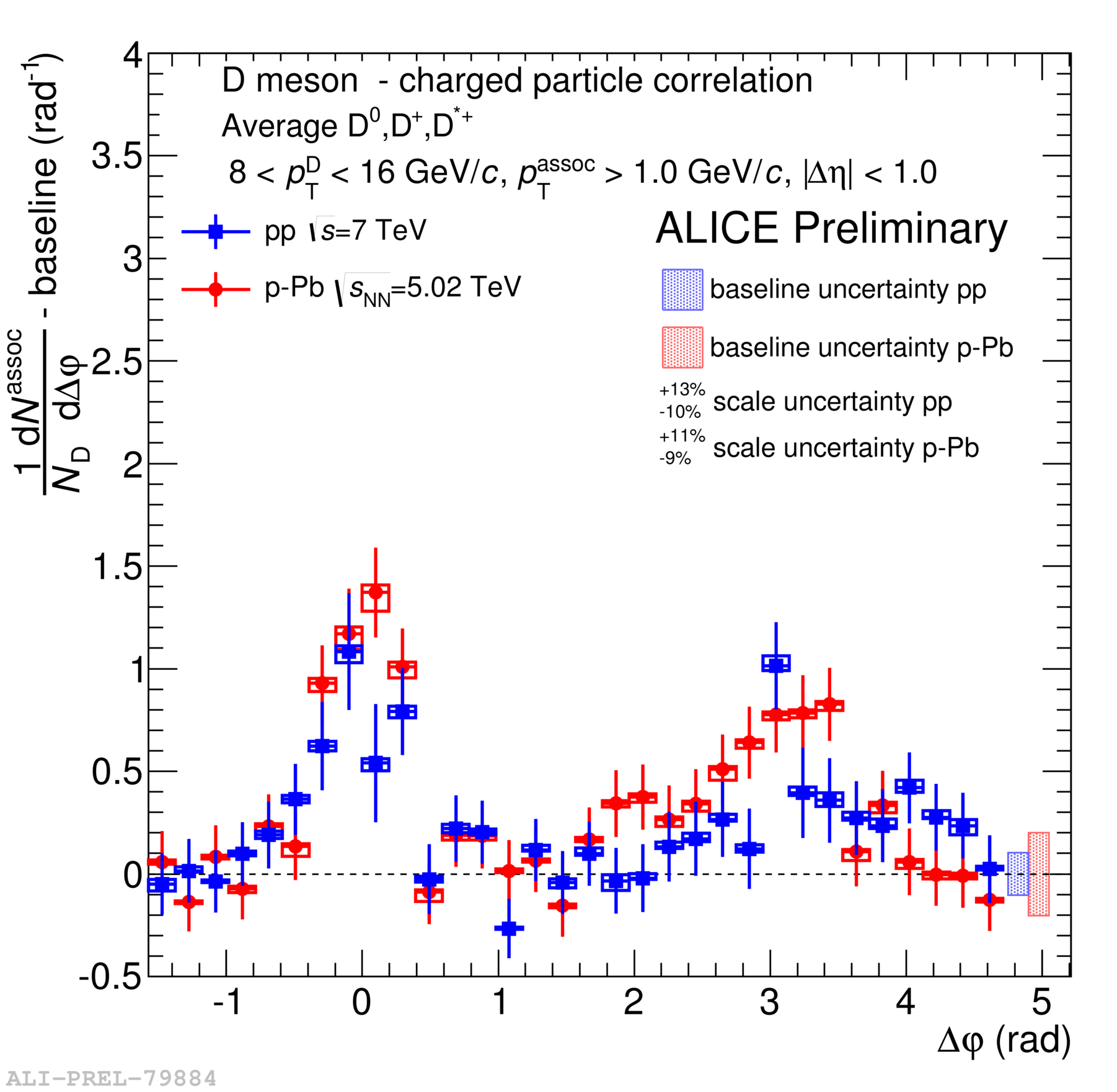}}}$
\end{minipage}
\caption{Comparison of D-hadron azimuthal correlation distributions in pp and p-Pb collisions, after baseline subtraction, measured with ALICE in two different kinematical ranges (weighted average of D$^0$, D$^+$ and D$^{\ast +}$ measurements). Statistical and uncorrelated systematic uncertainties are shown as error bars and boxes, respectively.}
\label{fig:RisData}
\end{figure}

\begin{figure}[!]
\centering
\begin{minipage}{\linewidth}
  \centering
  $\vcenter{\hbox{\includegraphics[width=.41\linewidth]{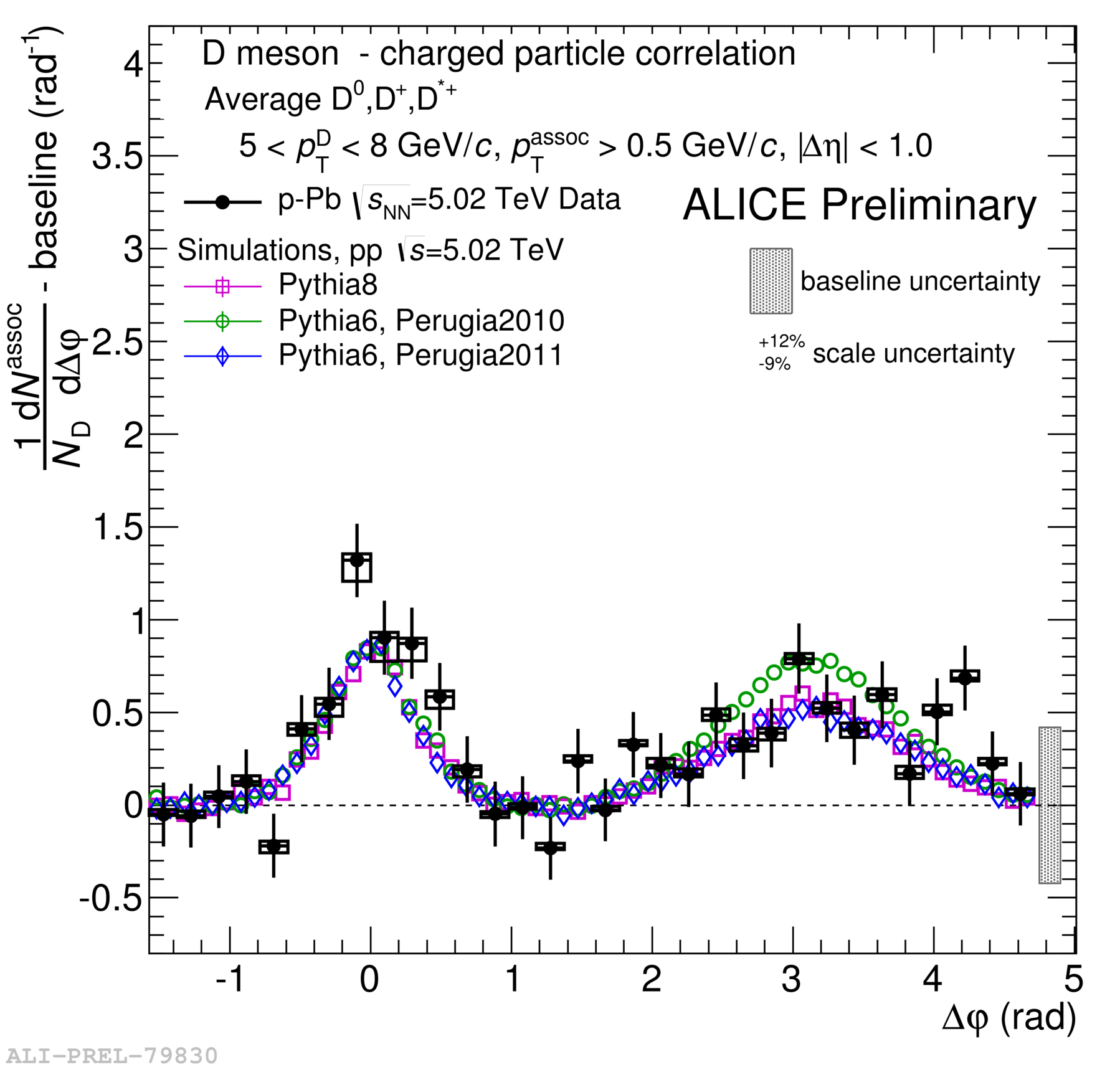}}}$ \hspace{0.5cm}
  $\vcenter{\hbox{\includegraphics[width=.41\linewidth]{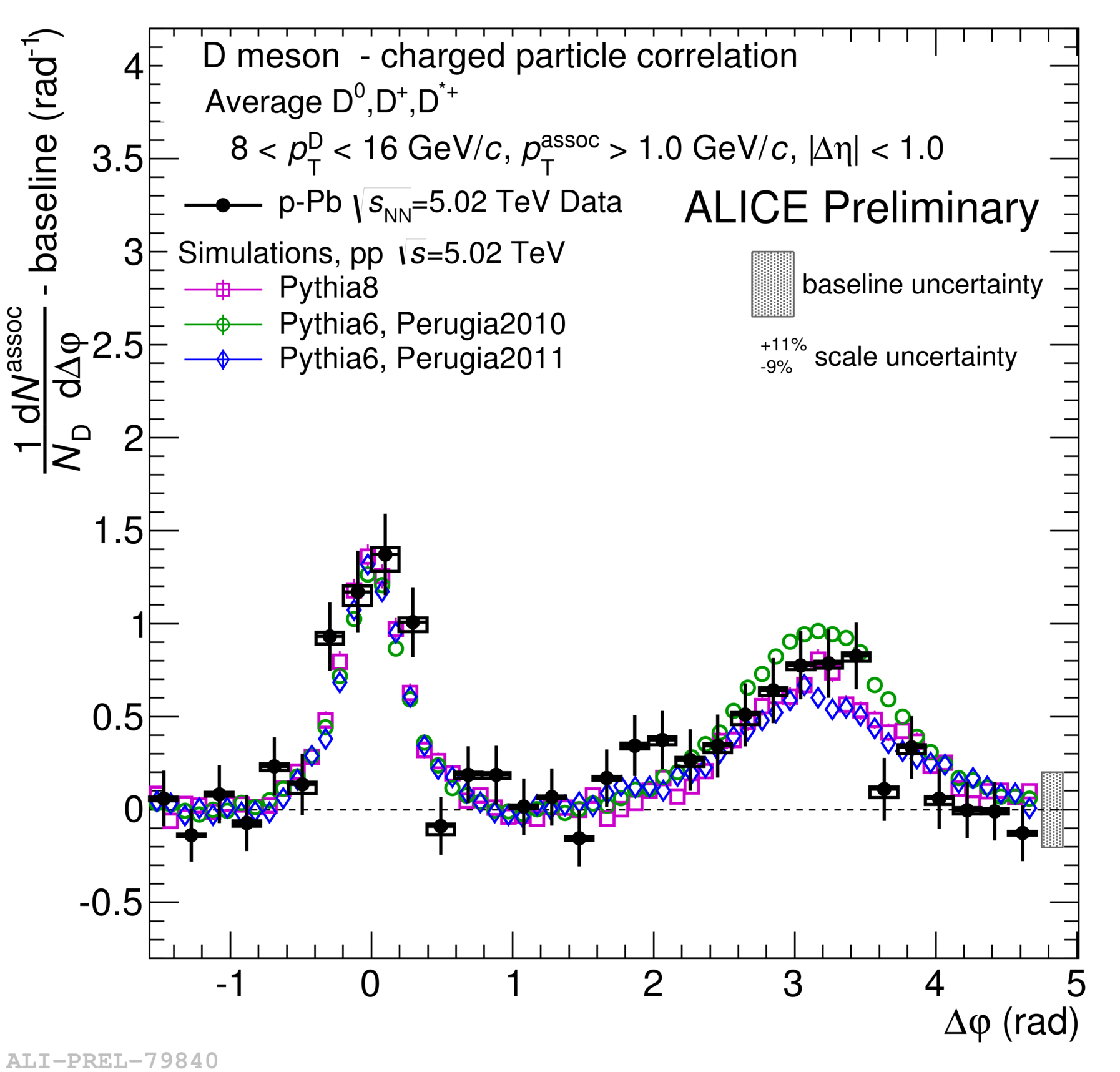}}}$
\end{minipage}
\caption{Comparison of D-hadron azimuthal correlation distributions in p-Pb collisions and predictions from different tunes of the PYTHIA generator, for pp collisions, after baseline subtraction, for two different kinematical ranges. Statistical and uncorrelated systematic uncertainties are shown as error bars and boxes, respectively.}
\label{fig:Pythia}       
\end{figure}

\begin{figure}[!]
\centering
\begin{minipage}{\linewidth}
  \centering
  $\vcenter{\hbox{\includegraphics[width=.41\linewidth]{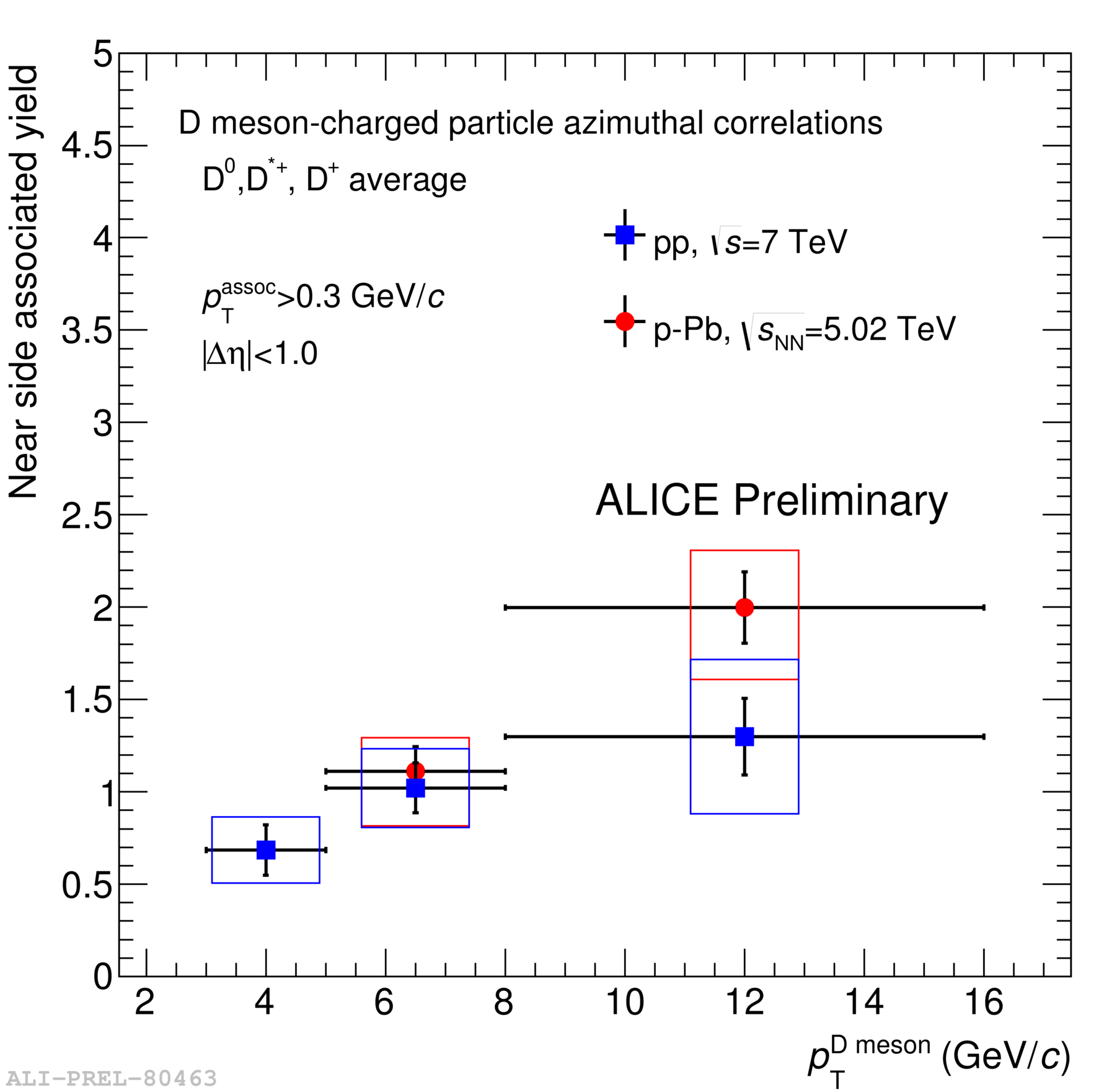}}}$ \hspace{0.5cm}
  $\vcenter{\hbox{\includegraphics[width=.41\linewidth]{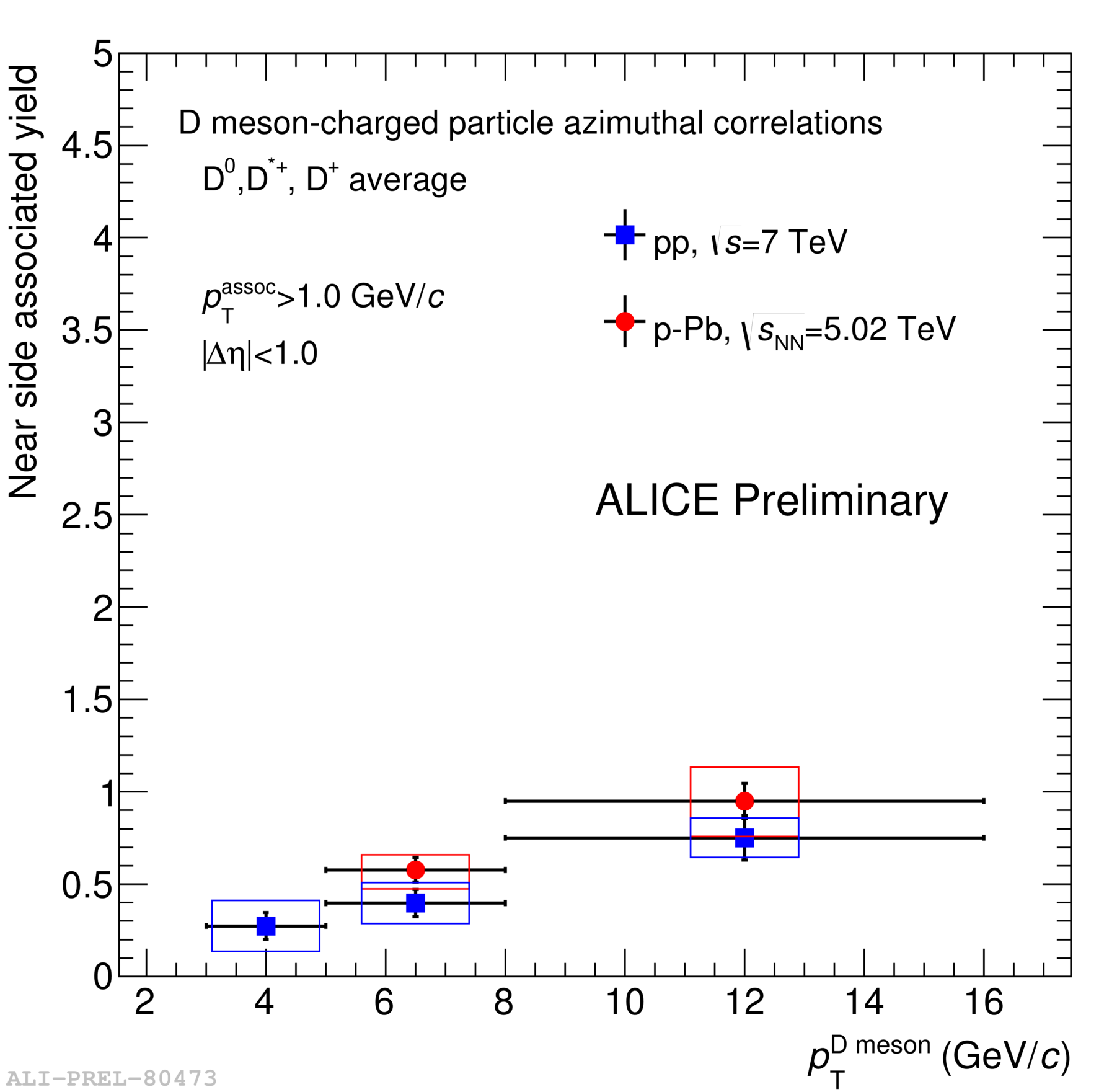}}}$
\end{minipage}
\caption{Comparison of near side associated yields extracted in in pp and p-Pb collisions as a function of the D-meson $p_{\rm T}$ for $p_{\rm T}^{\rm assoc} > 0.3$ (left panel) and 1 (right panel) GeV/$c$. Statistical and uncorrelated systematic uncertainties are shown as error bars and boxes, respectively.}
\label{fig:NSY}       
\end{figure}

Figure~\ref{fig:RisData} shows a comparison of D-hadron correlations in pp and p-Pb collisions, after the subtraction of the baseline, for $5< p_{\rm T}^{\rm D} < 8$ GeV/$c$ and associated tracks with $p_{\rm T}^{\rm assoc} > 0.5$ GeV/$c$ (left) and for $8 < p_{\rm T}^{\rm D} < 16$ GeV/$c$ and $p_{\rm T}^{\rm assoc} > 1.0$ GeV/$c$ (right). In both systems, the near ($\Delta\varphi \sim 0$) and away side ($\Delta\varphi \sim \pi$) peaks are clearly visible, despite the fluctuations due to the limited statistics available, and the correlation pattern is similar for pp and p-Pb collisions. In Fig.~\ref{fig:Pythia} the baseline-subtracted D-hadron correlation distributions extracted from the p-Pb analysis are compared with the expectations obtained from PYTHIA simulations, at $\sqrt{s} = 5.02$ TeV but for pp collisions, exploiting different tunings of the simulation parameters. An overall compatibility within the uncertainty is found between data and the predictions from the three PYTHIA tunes.
A more quantitative comparison between pp and p-Pb results for the near side region can be performed by extracting the yields of the near side peaks from the fit function applied on the correlation distributions (where the near side peak is modeled with a Gaussian). The near side associated yields obtained from pp and p-Pb correlation distributions are shown in Fig.~\ref{fig:NSY} as a function of the D-meson $p_{\rm T}$ for associated tracks with $p_{\rm T}^{\rm assoc} > 0.3$ GeV/$c$ (left) and $p_{\rm T}^{\rm assoc} > 1.0$ GeV/$c$ (right). Apart from the $3 < p_{\rm T}^{\rm D} < 5$ GeV/$c$ interval, where p-Pb results are not available, the pp and p-Pb yields are compatible within the total uncertainties. Given the size of the current uncertainties, no firm conclusions on modifications of D-hadron correlations due to cold nuclear matter effects can be drawn.
\section{Perspectives for Pb-Pb collisions}
\label{sec:lead}
The D-hadron correlation analysis is not feasible on the collected statistics of Pb-Pb collisions, with the current performance of the ALICE detectors. The main limiting factors are the very low D-meson signal/background ratio and the large amount of tracks from the underlying event, uncorrelated to D mesons. These issues are reflected in substantial statistical fluctuations in the $\Delta\varphi$ distributions, induced by the background subtraction, which wash out any correlation structure possibly present.

After the ALICE upgrade (expected for 2018-2019)~\cite{bib:LoI, bib:UpgITS}, though, a dramatic improvement of ITS tracking and vertexing performance is expected, which will translate into an increase of the $S/B$ ratio by a factor up to 10 for the D$^0$-meson reconstruction (and similarly for the other mesons). Combined with the increase of a factor $\approx$100 in the expected Pb-Pb statistics, this will allow the study of D-hadron correlations also in this system.

\begin{figure}[!]
\centering
\begin{minipage}{\linewidth}
  \centering
  $\vcenter{\hbox{\includegraphics[width=.42\linewidth]{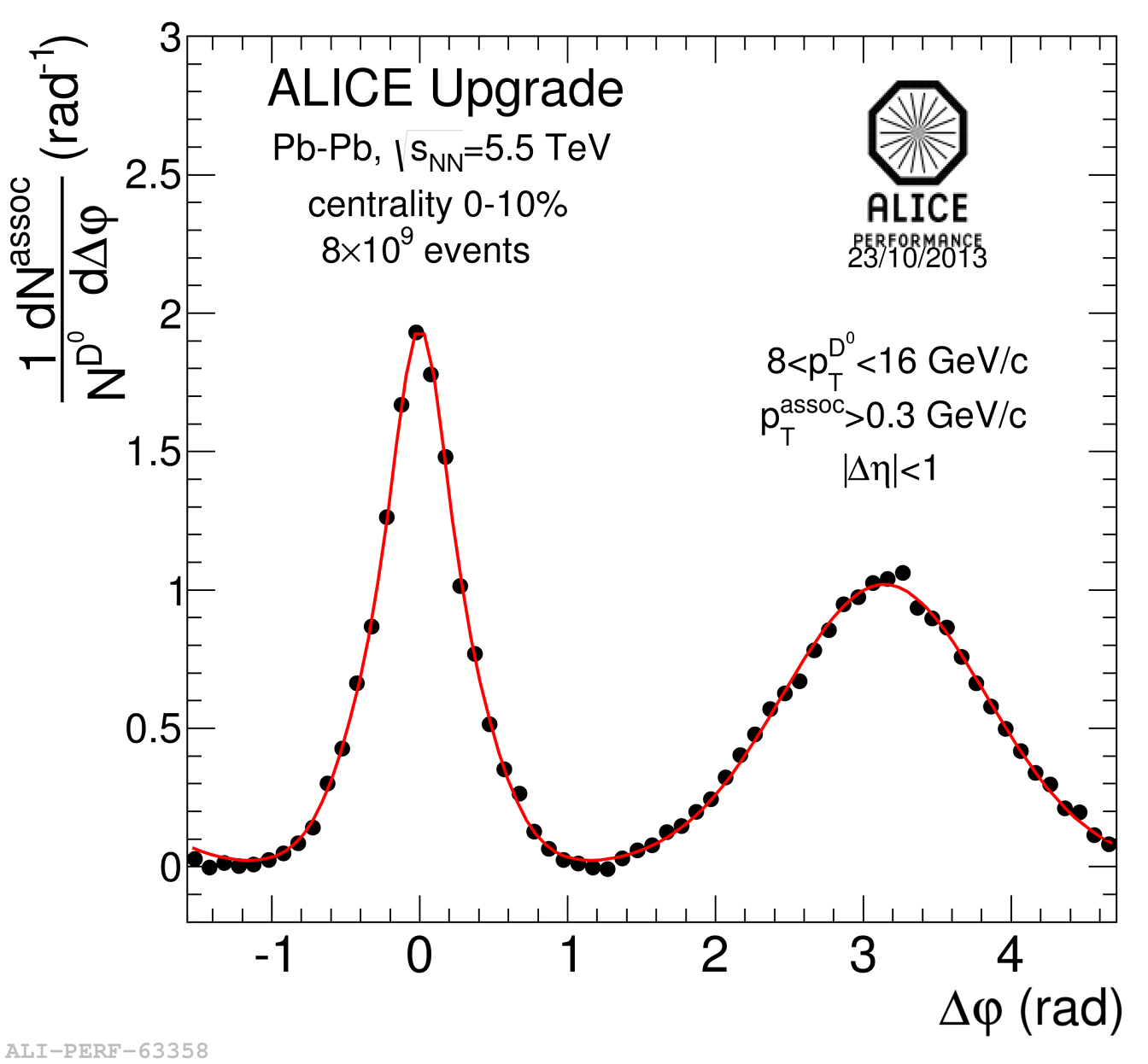}}}$ \hspace{0.5cm}
  $\vcenter{\hbox{\includegraphics[width=.41\linewidth]{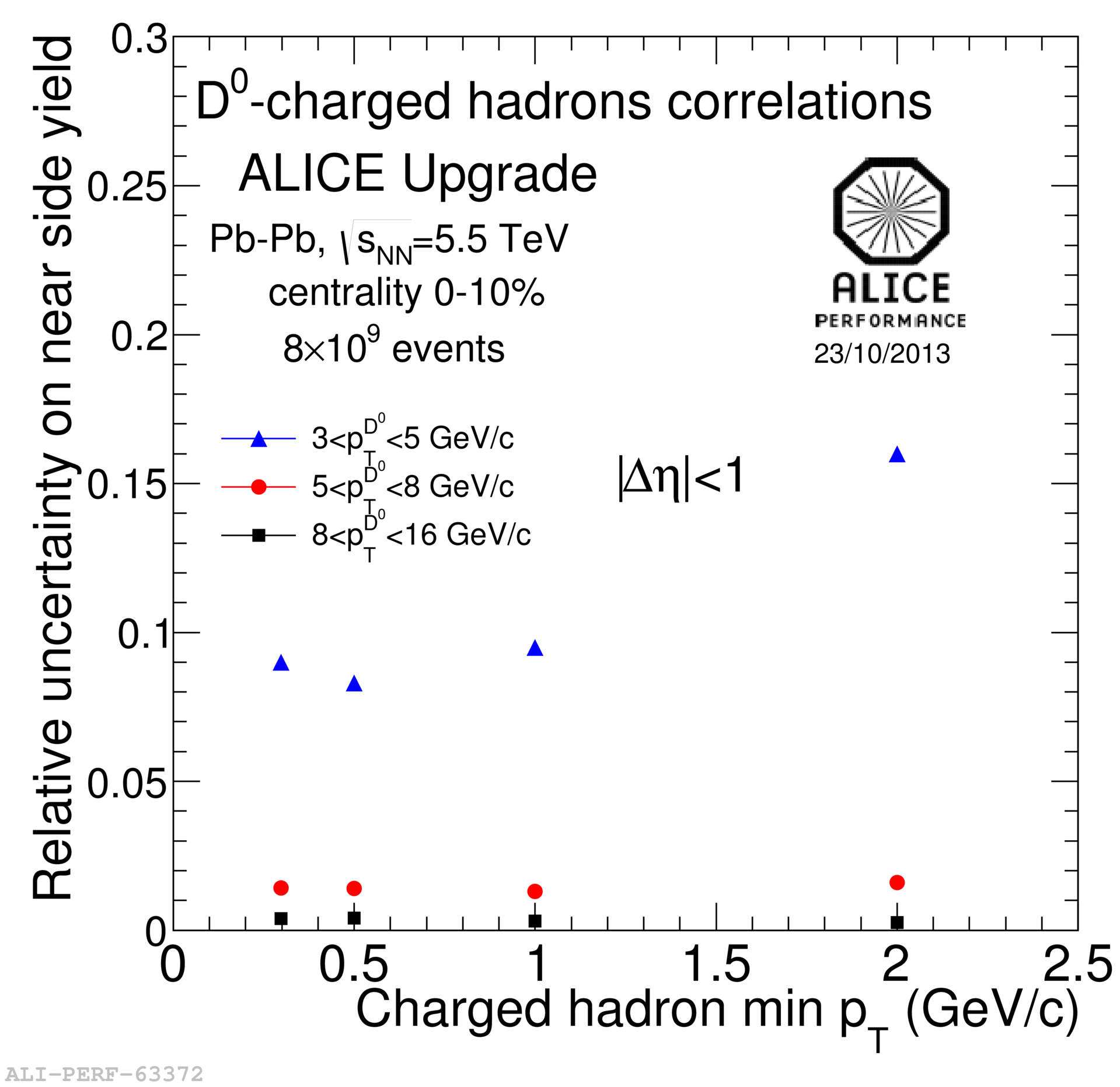}}}$
\end{minipage}
\caption{Left: estimated D$^0$-hadron azimuthal correlation distribution in 0-10\% central Pb-Pb collisions from a Monte Carlo simulation with the upgraded ALICE detectors for the $8 < p_{\rm T}^{{\rm D}^0} < 16$ GeV/$c$ range with $p_{\rm T}^{\rm assoc} > 0.3$ GeV/$c$ after the subtraction of the baseline. Right: estimates for the statistical uncertainty of the near side yield from the analysis of the expected Pb-Pb statistics after the ALICE upgrade for different kinematical ranges.}
\label{fig:pbpbUpg}       
\end{figure}

A simulation of the analysis performance on central (0-10\%) Pb-Pb collisions was carried out using a template of correlation distributions from PYTHIA. Figure~\ref{fig:pbpbUpg} shows an example of the D$^0$-hadron correlation pattern (left), where statistical fluctuations have a negligible effect. This leads to a very low statistical uncertainty on the evaluation of the near side yield, as shown in the right panel of Fig.~\ref{fig:pbpbUpg}, below 1\% for $p_{\rm T}^{{\rm D}^0} > 8$ GeV/$c$ and about 10\% for $3 < p_{\rm T}^{{\rm D}^0} < 5$ GeV/$c$. A substantial reduction of the uncertainties is expected also for pp and p-Pb collision systems. Consequently, after the ALICE upgrade it should be possible to compare the results in the three collision systems and to quantify the effects of the charm quark energy loss in a QGP medium.
\section{Conclusions}
\label{sec:lead}

Measurements of D-hadron angular correlations in pp collisions at $\sqrt{s} = 7$ TeV and in p-Pb collisions at $\sqrt{s_{\rm NN}} = 5.02$ TeV in different kinematical ranges have been presented. From the comparison of the results in the two collision systems, compatibility within the uncertainties is found both for the correlation shapes and for the near side associated yields. Consequently, no evident effects on the charm fragmentation and hadronization due to cold nuclear matter can be claimed.

While the analysis cannot be currently performed in Pb-Pb collisions, it will become feasible also in this collision system after the ALICE upgrade (2018-2019), when the tracking and vertexing performance of the experiment will be greatly enhanced.

\end{document}